\documentclass[prl,superscriptaddress,twocolumn,showpacs,amsmath,amssymb,floatfix]{revtex4-1}
\usepackage{graphicx}
\newcommand{\dto}{$\text{Dy}_2\text{Ti}_2\text{O}_7$}
\newcommand{\yto}{$\text{Y}_2\text{Ti}_2\text{O}_7$}

\newcommand{\ruo}{$\text{Ru}\text{O}_2$}
\newcommand{\figref}[1]{figure \ref{#1}}

\newcommand{\zz}{\mbox{2in-2out}}
\newcommand{\ed}{\mbox{1in-3out}}
\newcommand{\de}{\mbox{3in-1out}}
\newcommand{\Dy}{$\text{Dy}^{3+}$}

\begin{document}
\title{Thermal Conductivity and Specific Heat of the Spin-Ice Compound \dto:\\
       Experimental Evidence for Monopole Heat Transport}
\author{G.~Kolland}
\author{O.~Breunig}
\author{M.~Valldor}
\author{M.~Hiertz}
\author{J.~Frielingsdorf}
\author{T.~Lorenz}\email[E-mail: ]{tl@ph2.uni-koeln.de}
\affiliation{{\protect II.}\ Physikalisches Institut, Universit\"at zu K\"oln, Z\"ulpicher Str.\ 77, 50937 K\"oln, Germany}

\begin{abstract}
Elementary excitations in the spin-ice compound \dto\ can be described as magnetic monopoles propagating independently within
the pyrochlore lattice formed by magnetic Dy~ions. We studied the magnetic-field dependence of the thermal
conductivity $\kappa(B)$ for $B\,||\,[001]$ and observe clear evidence for magnetic heat transport originating from the monopole excitations.
The magnetic contribution $\kappa_\text{mag}$ is strongly field-dependent and correlates with the magnetization $M(B)$.
The diffusion coefficient obtained from the ratio of $\kappa_\text{mag}$ and the magnetic specific heat is strongly enhanced below \mbox{1 K}
indicating a high mobility of the monopole excitations in the spin-ice state.
\end{abstract}


\date{\today}
\maketitle

The recent prediction of magnetic monopoles in the spin-ice compounds has attracted a lot of interest \cite{Castelnovo2008,Morris2009,Giblin2011,Castelnovo2011,PhysRevLett.105.267205,Kadowaki2009,Jaubert2011,Bramwell2009,Blundell2012,Yaraskavitch2012}.
Spin ice is a geometrically frustrated spin system, which is realized in \dto\ by a sublattice of corner-sharing \Dy\ tetrahedra.
Due to a strong Ising anisotropy, the magnetic moments of \Dy\ point either in or out of a tetrahedron.
The  magnetic dipole energy is minimized, when two spins point in and two out of a tetrahedron (\zz),
what is realized by 6 out of $2^4=16$ possible configurations for a single tetrahedron.
In this respect, the spin orientation of \Dy\ corresponds to the hydrogen displacement in water ice \cite{Nagle1966}
and the ground state is highly degenerate with a residual entropy $ S_0=Nk_B/2\ln(3/2)$ for \mbox{$T\to0\,\text K$}
\cite{Ramirez1999,Bramwell2001,Hiroi2003,Sakakibara2003}.
Excited states can be created by flipping one spin, resulting in two adjacent tetrahedra with configurations \de\ and \ed, respectively.
In zero magnetic field, such a dipole excitation can fractionalize into two individual excitations, a monopole (\de) and an anti-monopole (\ed),
which can propagate independently. This can be visualized by flipping, \emph{e.g.},
another in-pointing spin of the \de\ tetrahedron such that it relaxes back to (another) \zz\ ground state configuration,
while the \de\ state has moved to a neighboring tetrahedron. 

\begin{figure}[t] \includegraphics[width=0.9 \linewidth]{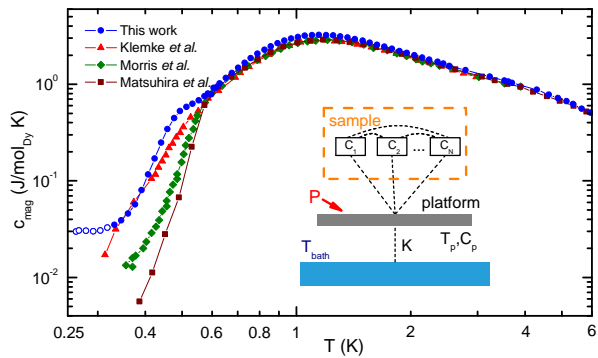}
  \caption{\label{c_mit_methode}(color online). Comparison of our specific heat data with literature data \cite{Klemke2011,Morris2009,Matsuhira2002}. The low-temperature plateau of our data (open symbols) probably arises from nuclear
  contributions of $^{161}\text{Dy}$ and $^{163}\text{Dy}$.
    Inset: Schematic illustration of the method to measure the specific heat.}
\end{figure}

The model of magnetic monopoles has been widely used to describe many experimental observations of \dto
\cite{Castelnovo2008,Morris2009,Giblin2011,Castelnovo2011,PhysRevLett.105.267205,Kadowaki2009,Jaubert2011,Bramwell2009,Blundell2012,Yaraskavitch2012}.
Nevertheless, there are basic properties of the spin-ice materials, which are far from being understood.
For example, the  specific heat $c_p$ of \dto\ has a pronounced maximum around \mbox{1.2 K}
resulting from the magnetic excitations and the corresponding entropy is close to the expected \mbox{$S_0=1.68\,\text{J}\,\text{K}^{-1}\,\text{mol}_\text{Dy}^{-1}$}
\cite{Matsuhira2002,Hiroi2003,Morris2009,Klemke2011}. Below \mbox{$\sim 600\,\text{mK}$}, however,
$c_p(T)$ data published by several groups differ by almost an order of magnitude, see figure \ref{c_mit_methode} \cite{Klemke2011,Morris2009,Matsuhira2002,residualentropy}.
Very recently it became clear that in this low-temperature regime the magnetic subsystem of \dto\ enters a region of very slow dynamics
with relaxation processes that may extend over extremely long time scales \cite{Klemke2011,Yaraskavitch2012}.
Another open issue is the dynamics of the magnetic monopoles.
The possible observation of a monopole current in an external magnetic field is currently under strong debate~\cite{Bramwell2009,dunsiger2011,Blundell2012}.
In this context, it is also unclear whether the magnetic monopoles contribute to the energy transport
and how the monopoles interact with each other and with the phonon excitations.
A suitable probe to study these issues are measurements of the thermal conductivity $\kappa$, which are the main subject of this report.
Only one experimental study about the heat conductivity of \dto\ has been published so far  \cite{Klemke2011}.
This previous work focuses on the anomalously enhanced relaxation times of \dto\ in the low-temperature region,
which are analyzed by assuming that the thermal conductivity is of purely phononic origin.
In this report, we present a detailed study of the magnetic-field dependent thermal conductivity $\kappa(B)$,
which yields clear evidence that up to almost 50\% of the low-temperature heat transport of \dto\ is of magnetic origin and our analysis 
suggests that this is a consequence of the high mobility of magnetic monopoles in zero field. 

For the actual study, large single crystals of \dto\ and \yto\ were grown by the floating-zone technique.
The crystals were oriented in a Laue camera and samples of typical sizes of several \mbox{$\text{mm}^3$} were cut.
The thermal conductivity was measured by the standard steady-state method from $\simeq 0.3$ to 300~K.
The temperature gradient was produced by a heater at one end of the sample and measured by two matched \ruo\ thermometers for $T<15$~K in magnetic fields up to 1~T.
In a separate run, the measurements were extended up to 300 K by using AuFe(0.07\%)-Chromel thermocouples to detect the temperature gradient.
The magnetic field was applied along \mbox{$[001]$}, perpendicular to the heat current along the \mbox{$[110]$} direction of a sample of
\mbox{$1 \times 1\times 3\,\text{mm}^3$} with the long edge parallel to [110].
For this sample geometry, demagnetization corrections have to be taken into account.
The magnetization, needed for the correction, has been measured with a home-built Faraday magnetometer on a thin sample of
\mbox{$0.3 \times 2 \times 4\,\text{mm}^3$} with the long axis $\| [001]$ to minimize demagnetization effects.

The specific heat was measured in a home-built calorimeter using a quasi-adiabatic heat-pulse method. A sample of \mbox{21 mg} was fixed to the platform by a small amount of grease. The addenda was measured in a separate run and was subtracted. It has been shown recently \cite{Klemke2011}, that below 600~mK the temperature relaxation of \dto\ contains multiple time constants, which can be understood by \dto\ consisting of subsystems weakly coupled to each other and to the platform. Standard methods to measure $c_p$ do not account for such sample-internal dynamics, and thus provide too small values \cite{Klemke2011}. Our approach to measure $c_p$ below \mbox{600 mK}, slightly differs from that of \cite{Klemke2011}. As the sample is at low temperature and in high vacuum we can safely assume that the subsystems $C_i$ are not  directly linked to the bath, but only to each other and to the platform (inset of \mbox{figure \ref{c_mit_methode}}), whose temperature $T_\text{P}$ is measured as a function of time. Starting from equilibrium (\mbox{$T_\text{sample}=T_\text P=T_\text{bath}$}), a constant heating power $P$ is applied to the platform until saturation is reached (\mbox{$T_\text{sample}\simeq T_\text{P} = T_\text{bath}+\Delta T$} \cite{cp-saturation}). The heat \mbox{$\Delta Q$} stored in the sample is then calculated from the total dissipated heat by subtracting the numerically obtained heat flown from the platform via $K$ to the bath. Above \mbox{600 mK}, this method yields heat capacities consistent with those obtained by conventional single-relaxation methods. At lower temperatures, however, this technique yields enhanced $c_p(T)$ data, similar to those of \cite{Klemke2011}, but without any {\it a-priori} assumptions about the subsystems or their respective couplings. \mbox{Figure \ref{c_mit_methode}} compares our results with literature data \cite{Klemke2011, Morris2009,Matsuhira2002}, which 
all agree with each other within $10\%$ above  $600\,\text{mK}$. For lower temperatures, however, the standard techniques  \cite{Morris2009,Matsuhira2002} result in significantly lower $c_p$ values than those obtained by the methods which explicitly account for the glassy behavior of different subsystems in \dto . In comparison to \cite{Klemke2011}, our data show an additional shoulder in $c_p(T)$ around 500~mK and a tendency towards saturation below 350~mK. We suspect that nuclear contributions to $c_p$ resulting from the isotopes $^{161}\text{Dy}$ and $^{163}\text{Dy}$ \cite{Anderson1969} may, at least partly, cause the anomalous low-temperature behavior of our data, which is not observed in \cite{Klemke2011}, where a $^{162}\text{Dy}$-enriched sample was studied. In order to clarify the origin of these deviations, further studies on different samples are necessary. For the following discussion, this uncertainty is of minor importance, because (i) the thermal conductivity $\kappa(B)$ has only been measured above 350~mK and (ii) the deviation between our $c_p(T)$ and the data of \cite{Klemke2011} around 500~mK will affect only a single point of the diffusion coefficient discussed below. Nevertheless, we emphasize that due to the glassy low-temperature behavior of \dto\ below 600~mK any quantitative comparison of experimental data with theoretical models should be treated with some caution. 

\begin{figure}[t]
  \includegraphics[width=0.9 \linewidth]{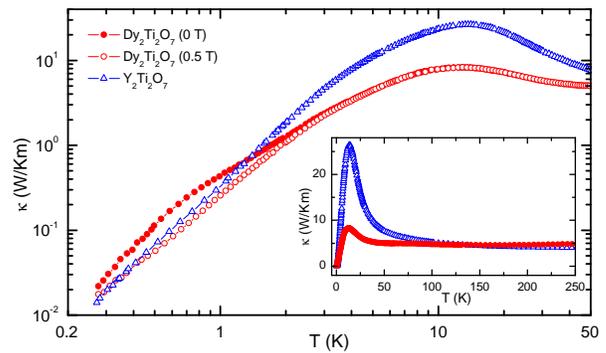}
  \caption{\label{vgl_dto_yto}(color online). Thermal conductivity of \dto\ for zero field and for \mbox{0.5 T} and the
    zero-field thermal conductivity of the non-magnetic \yto.
    Inset: Same data on linear scales up to \mbox{250 K}.}
\end{figure}

\begin{figure}[t]
  \includegraphics[width=0.95 \linewidth]{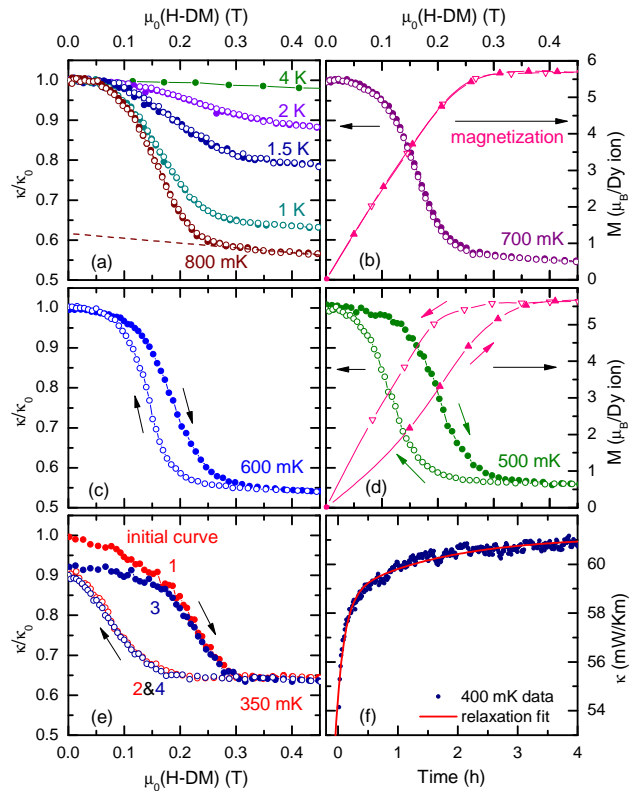}
  \caption{\label{kvb_B100}(color online). Field dependence of \mbox{$\kappa(B)/\kappa(0\,\text T)$} for \mbox{$B\,||\,[001]$}.
    All curves were measured after zero-field cooling.
    The fit of the linear decrease above \mbox{$\sim0.3\,\text T$} is shown exemplarily for 800~mK.
    Panels (b) and (d) also contain the magnetization data $M(B)$ ($\bigtriangleup$),
    and (f) displays the time dependent relaxation $\kappa(0\text T,t)$ back to the initial zero-field value for 400~mK.}
\end{figure}

The thermal conductivity \mbox{$\kappa(T)$} of \dto\ was measured at zero and in an external magnetic field of 1~T (\mbox{figure \ref{vgl_dto_yto}}), but as mentioned above the external field was corrected by a considerable demagnetization field of up to \mbox{0.5 T}.
Above \mbox{$4\,\text K$}, there is almost no field-dependence of \mbox{$\kappa(T)$}, while $\kappa$ is
significantly suppressed by a magnetic field for lower temperatures.
In order to obtain information about the phononic contribution of \mbox{$\kappa(T)$}, we also studied the iso-structural, but non-magnetic \yto.
As can be seen in the inset of \figref{vgl_dto_yto}, \mbox{$\kappa(T)$} of \dto\ and of \yto\ are very similar above 
100~K, but at lower temperature  \mbox{$\kappa(T)$} of \yto\ is significantly higher than that of \dto. Probably, this difference is related to an anomaly observed at \mbox{$110\,\text K$} in Raman spectroscopy data that indicates a structural instability in \dto\ which is absent in the non-magnetic homologue $\text{Lu}_2\text{Ti}_2\text{O}_7$ \cite{Saha2008,Kamaraju2011}. In addition, $\kappa$ of \dto\ can be reduced by phonon scattering via crystal-field excitations of the partly filled $4f$ shell of Dy. Considering the low-temperature range, it turns out that  \mbox{$\kappa(T)$} of \yto\ follows a power-law \mbox{$\kappa(T) \propto T^{2.4} $} below \mbox{3 K} and a similar behavior (\mbox{$\kappa(T) \propto T^{2.2} $}) is present for the \mbox{$\kappa(T)$} data of \dto\ in a magnetic field of \mbox{0.5 T}. In contrast, the zero-field \mbox{$\kappa(T)$} of \dto\ shows a clear shoulder around 1~K. This qualitative difference suggests the existence of an additional magnetic contribution $\kappa_\text{mag}$, which appears in the zero-field data on top of the phononic background, that is 
\begin{equation}
  \kappa = \kappa_\text{ph} + \kappa_\text{mag}\, ,
\end{equation}
where $\kappa_\text{ph}$ is essentially represented by the \mbox{$\kappa(T)$} data measured in 0.5~T.
This conclusion is confirmed by measurements of the field-dependent \mbox{$\kappa(B)$}
at constant temperatures. Again, demagnetization has been taken into account to rescale the magnetic field.
Figure~\ref{kvb_B100} displays the relative change \mbox{$\kappa/\kappa_0$}
for different constant temperatures. Below 2~K, we find a step-like decrease of $\kappa(B)$ around 0.2~T,
which systematically sharpens on decreasing temperature. As is shown exemplarily for \mbox{800 mK} in \mbox{figure \ref{kvb_B100}a},
this step is followed by a weak, essentially linear decrease towards higher fields.
The relative reduction $\kappa/\kappa_0$ has a maximum around \mbox{600 mK} and vanishes practically above \mbox{4 K}.
The decrease of $\kappa(B)$ correlates with the increase of \mbox{$M(B)$},
shown exemplarily for \mbox{700 mK} and \mbox{500 mK} in figure \ref{kvb_B100}b and \ref{kvb_B100}d, respectively.
The step-like change of \mbox{$\kappa/\kappa_0$} exhibits a clear hysteresis below \mbox{$700\,\text{mK}$} (figure \ref{kvb_B100}c-e),
with different critical fields depending on the field-sweep direction.

Finally, an additional feature appears in \mbox{$\kappa(B)$} below \mbox{$500\,\text{mK}$} (\mbox{figure \ref{kvb_B100}e}):
Measuring an initial curve (1) after zero-field cooling and subsequently decreasing the magnetic field back to zero (curve~2), $\kappa(0\,\text T)$ only recovers about 90\% of its original zero-field value. Repeating the field sweeps from this new starting point results in  \mbox{$\kappa(B)$} curves (3) and (4) with coinciding endpoints, where curves (2) and (4) perfectly match each other. The reduced zero-field values $\kappa(0\,\text T)$ slowly relax back to the respective  zero-field-cooled values, as is shown in Figure~\ref{kvb_B100}f for \mbox{$400\,\text{mK}$}.
In order to describe this relaxation process at least two relaxation times are needed, that is
\begin{equation}
  \kappa(t) = a_0+a_1(1-\text e^{-t/\tau_1})+a_2(1-\text e^{-t/\tau_2}) \, .
\end{equation}
The fit of figure \ref{kvb_B100}f yields large relaxation times \mbox{$\tau_1 \simeq 8\,\text{min}$} and \mbox{$\tau_2 \simeq 100\,\text{min}$}. As every data point of $\kappa(B)$ in figure \ref{kvb_B100}a-e requires several minutes of temperature stabilization, the different zero-field values (figure~\ref{kvb_B100}e)
originate from the extremely slow relaxation $\tau_2$ \cite{slowrelax}. This very slow relaxation only occurs after a field sweep below \mbox{$500\,\text{mK}$} but not after cooling at zero field.
Such a slow glass-like low-temperature behavior of \dto\ is in agreement with various recent reports \cite{PhysRevLett.105.267205,Klemke2011,Erfanifam2011,Aoki2004}.

A straightforward qualitative interpretation of the observed magnetic-field dependent $\kappa(B)$ is obtained by assuming that the magnetic contribution $\kappa_\text{mag}$ results from a heat transport by magnetic monopoles. Due to the degenerate zero-field ground state (\zz), a monopole excitation (\de\ or \ed) can easily propagate by single spin flips. When a monopole excitation passes through a \zz\ tetrahedron, two subsequent single spin flips are needed, which change the initial \zz\ configuration of this tetrahedron to another \zz\ state. As in zero magnetic field all \zz\ configurations are degenerate, the magnetic monopoles have a large mobility. This drastically changes in a magnetic field along [001], which lifts the degeneracy of the different \zz\ configurations and thus the monopole mobility is suppressed. In addition, the correlation between increasing magnetization and decreasing heat conductivity is explained. The magnetic field determines, which \zz\ state is the field-induced ground state, and for not too small fields (\mbox{$\gtrsim 0.1\,\text T$}), the magnetization is proportional to the population of this particular \zz\ state, whereas the monopole mobility systematically decreases the more this \zz\ state is populated. With increasing $B||[001]$, the monopole excitation energy also increases weakly, \emph{i.e.} the monopole density decreases, but we expect this effect to be of minor importance, because the decreasing monopole mobility due to the field-induced lifting of the ground-state degeneracy will be the dominating effect. For magnetic fields above the step-like decrease of \mbox{$\kappa(B)$}, the heat transport is purely phononic and in order to extract the zero-field phononic contribution, we extrapolate the weak linear decrease of \mbox{$\kappa(B)$} back to \mbox{$0\,\text{T}$} (\mbox{\figref{kvb_B100}a}). The extrapolation uncertainties determine the error bars of $\kappa_\text{mag}$.
\begin{figure}[t]
  \includegraphics[width=0.9 \linewidth]{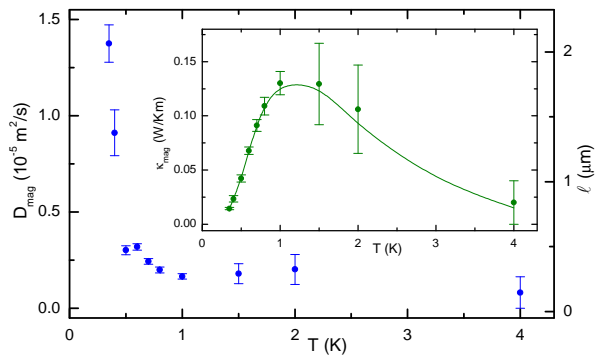}
  \caption{\label{D_mag}(color online). Magnetic contribution $\kappa_\text{mag}$ of the zero-field heat transport (inset; the line is to guide the eye) and the diffusion coefficient $D_\text{mag}=\kappa_\text{mag}/c_\text{mag}$,
  together with the estimated mean free path $\ell$.
}
\end{figure}
The resulting zero-field magnetic contribution $\kappa_\text{mag}$ has a  pronounced  maximum at \mbox{$\sim\,1.2\,\text K$} (inset of \mbox{\figref{D_mag}}). The maximum of $\kappa_\text{mag}$ is located close to that of $c_\text{mag}$, but on further increasing temperature $\kappa_\text{mag}$ rapidly decreases and practically vanishes above \mbox{$\sim\,4\,\text K$}. The main panel of \mbox{\figref{D_mag}} displays the corresponding diffusion coefficient,
\begin{equation}
   D_\text{mag} = \frac {\kappa_\text{mag}}{c_\text{mag}} \, .
  \label{eq_diffusion}
\end{equation}
Above 1~K, $D_\text{mag}$ is almost temperature independent but strongly increases towards lower temperatures. This fits to the qualitative expectation of a high monopole mobility by single spin flips on the degenerate ground state. At low temperatures, the monopole excitations are highly dilute resulting in a large mean free path and hence a large diffusion coefficient. With increasing temperature, the number of monopoles increases and thus the diffusion coefficient is expected to decrease. In the simplified model of single spin flips, this follows from the fact that a monopole can pass an already excited tetrahedron, \emph{i.e.} another (anti-)monopole, only via simultaneous spin flips. 
Within kinetic gas theory, the diffusion coefficient is related via $D = v\ell/3$ with the mean velocity $v$ and the mean free path $\ell$ of the particles, but this relation cannot be directly applied to the actual monopole gas in \dto\ for various reasons. First of all, the number of monopoles is not conserved and, moreover, neither the (average) velocity of monopoles nor their interaction with each other or with the phonon excitations is well understood up to now. Quite recently, an  expression for the monopole mobility in a (magnetic or electric) field has been derived~\cite{Castelnovo2011}, but it is unclear whether this result can be related to our experimental $D_\text{mag}$ arising from a finite temperature gradient in zero field. Despite these uncertainties, we give a rough estimate of a mean-free path by assuming a monopole velocity \mbox{$v = a_\text d\nu \approx 20\,\text m/\text s$}. Here, \mbox{$a_\text d = 4.34\,\text{\AA}$} denotes the distance of
neighboring tetrahedra and $\nu$ is the rate of single spin flips, which we roughly estimate by the monopole excitation energy \mbox{$E_\text m \approx 2.2\,\text K$} \cite{Castelnovo2011} to \mbox{$\nu = E_\text{m}/h \approx 4.6\cdot 10^{10}\,\text s^{-1}$}. Assuming a temperature-independent $v$, \mbox{$\ell = 3D_\text{mag}/v$} yields a linear scaling from $D_\text{mag}(T)$ to  $\ell(T)$. As shown in \mbox{figure \ref{D_mag}} (right scale), within this estimate $\ell(T)$ reaches the $\mu\text m$ range, \emph{i.e.} $\sim 1000$ unit cells, for \mbox{$T < 500\,\text {mK}$}, which may be understood from the low monopole density.
However, even around \mbox{$\sim 2\,\text K$}, $\ell$ is still of the order $\sim100$ unit cells although almost every second tetrahedron is in an excited \de\ or \ed\ configuration. This suggests that at least towards higher temperatures, more complex hopping models have to be involved to describe $\kappa_\text{mag}(T)$.

In conclusion, we observe clear evidence for a considerable magnetic contribution $\kappa_\text{mag}$ to the heat transport in the spin-ice material \dto . At constant temperature, the magnetic-field dependent decrease of $\kappa(B)$ correlates with the increase of the magnetization $M(B)$, which measures the population of the particular \zz\ configuration becoming the field-induced non-degenerate ground state. This reveals that the complete suppression of $\kappa_\text{mag}$ in a magnetic field of about \mbox{0.5 T} arises from the loss of monopole mobility or, \emph{vice versa}, the large $\kappa_\text{mag}(\text{0 T})$ is a consequence of the zero-field ground state degeneracy.
Our data also reveal a strong increase of the relaxation times below about 600~mK.
Including specific heat data, we derive a strong increase of the diffusion coefficient below 1~K, which is most probably related to the fact that the monopole gas is highly dilute towards low temperature. In order to derive quantitative information about, \emph{e.g.}, the mean-free path or the velocities, theoretical models about the monopole dynamics are required.

We acknowledge fruitful discussions with L.~Fritz, M.~Gr\"u\-nin\-ger, K.~Kiefer, and B.~Klemke and financial 
support by the Deutsche Forschungsgemeinschaft via SFB 608.


\begin{thebibliography}{26}

\bibitem{Castelnovo2008}
 C.~Castelnovo, R.~Moessner, and S.~L.~Sondhi,
 Nature {\bf 451}, 42 (2008)

\bibitem{Morris2009}
 D.~J.~P.~Morris {\em et al.},
 Science \textbf{326}, 411 (2009)

\bibitem{Giblin2011}
 S.~R.~Giblin {\em et al.},
 Nat. Phys. \textbf{7}, 252 (2011)

\bibitem{Castelnovo2011}
 C.~Castelnovo, R.~Moessner,  and S.~L.~Sondhi,
 Phys. Rev. B \textbf{84}, 144435 (2011)

\bibitem{PhysRevLett.105.267205}
 D.~Slobinsky {\em et al.},
 Phys. Rev. Lett. \textbf{105}, 267205 (2010)

\bibitem{Kadowaki2009}
 H.~Kadowaki {\em et al.},
 J. Phys. Soc. Jpn. \textbf{78}, 103706 (2009)

\bibitem{Jaubert2011}
 L.~D.~C.~Jaubert and P.~C.~W.~Holdsworth,
 J. Phys.: Condens. Matter \textbf{23}, 164222 (2011)

\bibitem{Bramwell2009}
 S.~T.~Bramwell {\em et al.},
 Nature \textbf{461}, 956 (2009)

\bibitem{Blundell2012}
 S.~J.~Blundell,
 Phys. Rev. Lett. \textbf{108}, 147601 (2012)

\bibitem{Yaraskavitch2012}
 L.~Yaraskavitch {\em et al.},
 Phys. Rev. B \textbf{85}, 020410 (2012)

\bibitem{Nagle1966}
 J.~F.~Nagle,
 J. Math. Phys. \textbf{7}, 1484 (1966)

\bibitem{Ramirez1999}
 A.~P.~Ramirez {\em et al.},
 Nature \textbf{399}, 333 (1999)

\bibitem{Bramwell2001}
 S.~T.~Bramwell and M.~J.~Gingras,
 Science \textbf{294}, 1495 (2001)

\bibitem{Hiroi2003}
 Z.~Hiroi {\em et al.},
 J. Phys. Soc. Jpn. \textbf{72}, 411 (2003)

\bibitem{Sakakibara2003}
 T.~Sakakibara {\em et al.},
 Phys. Rev. Lett. \textbf{90}, 207205 (2003)

\bibitem{Matsuhira2002}
 K.~Matsuhira {\em et al.},
 J. Phys.: Condens. Matter \textbf{14}, L559 (2002)

\bibitem{Klemke2011}
 B.~Klemke {\em et al.},
 J. Low Temp. Phys. \textbf{163}, 345 (2011)

\bibitem{residualentropy}
 These differences hardly change the entropy $\int c/T\,\text dT$.

\bibitem{dunsiger2011}
 S.~Dunsiger {\em et al.},
 Phys. Rev. Lett. \textbf{107}, 207207 (2011)

\bibitem{cp-saturation}
 Due to strongly increasing time scales below 500~mK and in order to avoid systematic errors induced by time-dependent experimental conditions, the heating curves $T_\text{P}(t)$ have to be cut off after suitable times, typically 5~--~10~minutes. Further relaxation can be estimated to increase $c_p$ by only a few percent. Thus, the presented data represent lower bounds for $c_p$.

\bibitem{Anderson1969}
 A.~Anderson {\em et al.},
 Phys. Rev. \textbf{183}, 546 (1969)

\bibitem{Saha2008}
 S.~Saha {\em et al.},
 Phys. Rev. B \textbf{78}, 214102 (2008)

\bibitem{Kamaraju2011}
 N.~Kamaraju {\em et al.},
 Phys. Rev. B \textbf{83}, 134104 (2011)

\bibitem{slowrelax}
 In order to capture $\tau_1$ in $\kappa(0\,{\text T}, 400\,{\text{mK}},t)$ weaker stabilization criteria were used.

\bibitem{Erfanifam2011}
 S.~Erfanifam {\em et al.},
 Phys. Rev. B \textbf{84}, 220404 (2011)

\bibitem{Aoki2004}
 H.~Aoki {\em et al.},
 J. Phys. Soc. Jpn. \textbf{73}, 2851 (2004)

\end{thebibliography}
\end{document}